\documentclass[nofootinbib,notitlepage,superscriptaddress,10pt,aps,pra,twocolumn]{revtex4-1}

\usepackage[utf8x]{inputenc}
\usepackage{amsmath}
\usepackage{amssymb}
\usepackage{bbm}
\usepackage{graphicx}
\usepackage[normalem]{ulem}

\newcommand{\ttt}{$\,\,$}

\begin{document}



\title{Planck-scale soccer-ball problem: a case of mistaken identity}

\author{Giovanni AMELINO-CAMELIA}
\affiliation{Dipartimento di Fisica, Universit\`a di Roma ``La Sapienza", P.le A. Moro 2, 00185 Roma, EU}
\affiliation{INFN, Sez.~Roma1, P.le A. Moro 2, 00185 Roma, EU}

\begin{abstract}
Over the last decade it has been found that nonlinear laws of composition of momenta are predicted by some alternative approaches to ``real" 4D quantum gravity, and by all  formulations of dimensionally-reduced (3D) quantum gravity coupled to matter. The possible relevance for
rather
different quantum-gravity models has motivated several studies,
but this interest is being tempered by concerns that a nonlinear law of addition of momenta might inevitably produce a pathological description of the total momentum of  a macroscopic body. I here show that such concerns are unjustified, finding that they are rooted in failure to appreciate the differences between two roles for laws composition of momentum in physics. Previous
results relied
exclusively on the role of a law of momentum composition in the description of spacetime locality. However, the notion of total
momentum of a multi-particle system is not a manifestation of locality,
but rather reflects translational invariance.
By working within an illustrative example of quantum spacetime I show explicitly that spacetime locality is indeed reflected in a nonlinear law of composition of momenta, but translational invariance  still results in an undeformed linear law of addition of momenta building up the  total momentum of a multi-particle system.
\end{abstract}

\maketitle

\section{Introduction}

An emerging  characteristic of quantum-gravity research over the last decade has been a gradual shift of focus  toward manifestations of the Planck scale on momentum space, particularly pronounced in some approaches to quantum gravity. For some research lines based on spacetime noncommutativity several momentum-space structures have been in focus, including the possibility of deformed laws of composition of momenta,
which shall be here
 of interest.
While deformed laws of composition of momenta are found to be
 inevitable in some approaches based on
spacetime noncommutativity (see, {\it e.g.},
Refs.\ttt\cite{majidLectNotes,lukiestar,alexstar,lukieAnnPhys,gaclukieIJMP,jurekphasespace}),
 the situation is less certain in
the loop-quantum-gravity approach. For ``real" 4D loop quantum gravity
the relevant issues are partly obscured by our present limited understanding of the semiclassical limit of that theory\ttt\cite{rovelliLRR08},
but some indirect arguments suggest that a nonlinear law of composition
of momenta might arise\ttt\cite{leeLOOPkodamaDSR}.
 And these arguments find further strength in results on  3D\ loop quantum gravity, where the simplifications afforded by that dimensionally-reduced model allow one to rigorously show that indeed the nonlinearities on momentum space are present (see, {\it e.g.}, Ref.\ttt\cite{noui3DLOOP}). Actually, evidence is growing that in all alternative formulations of 3D quantum gravity coupled to matter there are nonlinearities in momentum space, including nonlinear laws of composition of momenta  (see, {\it e.g.}, Ref.\ttt\cite{freidellivinePRL}).
Also noteworthy is the role played by nonlinearities on momentum space  in two recently-proposed approaches to the quantum-gravity problem: the one based on group field theory\ttt\cite{oritiMOMENTUMholonomy}
and the one based on the relative-locality framework\ttt\cite{prl}.

Due to the lack of experimental guidance a variety of approaches to quantum gravity are being developed, and in most cases the different approaches have very little in common. This of course endows with additional reasons of interest any result which is found to apply to more than one approach. Indeed, there has been
growing interest in the conceptual implications and possible phenomenological
implications\ttt\cite{gacLRR} of nonlinear laws on momentum space and particularly nonlinear
laws of composition of momenta. However, this interest is
being tempered by concerns that a nonlinear law of addition of momenta might inevitably produce a pathological description of the total momentum of  a macroscopic body\ttt\cite{lukiesoccer,maggioresoccer,gacsoccer,kowasoccer,girellisoccer,jacosoccer,sabisoccer,mignemisoccer,magpasoccer}
(also see Refs.\ttt\cite{gacfkssoccerRL,sabisoccerRL,gacfkssoccerRL2},
for a related discussion focused within the novel relative-locality
framework).
This issue has been often labelled as
the ``soccer-ball problem"\ttt\cite{gacsoccer}: the quantum-gravity
pictures lead one to expect nonlinearities of the law of composition
of momenta which are suppressed by the Planck scale ($\sim 10^{28} eV$)
and would be unobservably small for particles at energies we presently
can access, but in the analysis of a macroscopic body,
such as a soccer ball,
one might have to add up very many of
such minute nonlinearities, ultimately obtaining results in conflict with
observations\ttt\cite{lukiesoccer,maggioresoccer,gacsoccer,kowasoccer,girellisoccer,jacosoccer,sabisoccer,mignemisoccer,magpasoccer}.

If this so-called ``soccer-ball problem"
really was a scientific problem (a case of actual conflict with experimental data)
we could draw rather sharp conclusions about several areas of quantum-gravity
research. Perhaps most notably we should consider as ruled out
large branches of research on quantum-gravity based on spacetime noncommutativity and we should consider the whole effort of research on dimensionally-reduced 3D quantum gravity as completely unreliable in forming an intuition
for ``real" 4D quantum gravity.
However, I here show that previous discussions of this soccer-ball problem\ttt\cite{lukiesoccer,maggioresoccer,gacsoccer,kowasoccer,girellisoccer,jacosoccer,sabisoccer,mignemisoccer,magpasoccer,gacfkssoccerRL,sabisoccerRL,gacfkssoccerRL2}
  failed to appreciate the differences between two roles for laws of composition of momentum in physics. Previous
results supporting a nonlinear law of addition of momenta relied
exclusively on the role of a law of momentum composition in the description of spacetime locality. The notion of total
momentum of a multi-particle system is not a manifestation of locality,
but rather reflects translational invariance in interacting theories.
After being myself confused about these issues for quite
some time\ttt\cite{gacsoccer}
I feel I am now in position to articulate the needed discussion at a completely general level. However, 
considering the tone and content of the bulk of literature
that precedes this contribution of mine I find it is best to
opt here instead for a very explicit discussion, based on illustrative examples
of calculations performed within a specific
simple model affected by nonlinearities
for a law of composition of momenta. The model I focus on
has 2+1-dimensional
pure-spatial $\kappa$-Minkowski noncommutativity\ttt\cite{majidLectNotes,lukiestar,alexstar,lukieAnnPhys,gaclukieIJMP,jurekphasespace}, with
the time coordinate left unaffected by the
deformation and the two spatial coordinates, $x_1$ and $x_2$,
 governed by
\begin{equation}
[x_1,x_2]= i \ell x_1
\label{kappaspace}
\end{equation}
(with the deformation scale $\ell$  expected to be of the order of the inverse
of the Planck  scale).

In the next section I briefly review within this example
of quantum spacetime
previous arguments showing that
spacetime locality is reflected in a
nonlinear law of composition of momenta. Then Section III takes off from known results on translational invariance for $\kappa$-Minkowski noncommutative spacetimes and builds on those to achieve the first ever example of translationally-invariant interacting two-particle system in $\kappa$-Minkowski. This allows me to verify explicitly that the conserved charge associated
with that translational invariance (the total momentum of the two-particle system) adds linearly the momenta of the two particles involved.
Section IV offers some closing remarks.

\section{Soccer-ball problem and sum of momenta from locality}
The ingredients needed for seeing a nonlinear law of composition
of momenta emerging from noncommutativity of type (\ref{kappaspace}) are very simple. Essentially one needs only to rely on results
establishing that functions of coordinates governed by (\ref{kappaspace})
still admit a rather standard Fourier expansion (see, {\it e.g.},
Ref.\ttt\cite{majidLectNotes,lukiestar})
\begin{equation}
\Phi({ x})  = \int d^4 k \,\,
{\tilde \Phi}(k) ~ e^{ik_\mu{x}^\mu}
  \nonumber
\end{equation}
and that the notion of integration on such a
noncommutative space
preserves many  of the standard properties
including\ttt\cite{majidLectNotes,alexstar}
\begin{equation}
 \int d^4 { x} \,\,
 e^{ik_\mu{x}^\mu} = (2 \pi)^4 \delta^{(4)}(k) \, .
\end{equation}
It is a rather standard exercise for practitioners of
spacetime noncommutativity to use these  tools in order
to enforce locality within actions describing
 classical fields. For example, one might  want to introduce in
 the action the product of three (possibly identical, but in general
 different) fields, $\Phi$, $\Psi$, $\Upsilon$,
insisting on locality in the sense that the three fields
be evaluated ``at the same quantum point ${x}$", {\it i.e.} $\Phi({ x}) \, \Psi({ x})
  \,  \Upsilon(x)$. There is still no consensus on how
  one should formulate the more interesting quantum-field version
of such theories, and it remains unclear to which extent and in
which way our ordinary notion of locality
is generalized by the requirement of evaluating ``at the same
quantum point ${x}\,$" fields intervening in a product
such as $\Phi({ x}) \, \Psi({x})
  \,  \Upsilon({ x})$. Nonetheless for
  the classical-field case there is a sizable literature
  consistently adopting this prescription for locality. Important
  for my purposes here is the fact that, with such a prescription,
  locality inevitably leads to a nonlinear law of composition
  of momenta, as I show explicitly in the following
  example:
\begin{eqnarray}
&& \int d^4 {\hat x} \,
\,\Phi({\hat x}) ~\Psi({\hat x})
  ~  \Upsilon({\hat x}) =  \label{soccerball}\\
&& \,\,\,\,\,
= \int d^4 {\hat x}\int d^4 k \int d^4 p \int d^4 q \,
{\tilde \Phi}(k) \, {\tilde \Psi}(p)
  \, {\tilde \Upsilon}(q) \, e^{ik_\mu{\hat x}^\mu}
  e^{ip_\nu{\hat x}^\nu} e^{iq_\rho{\hat x}^\rho}
  \nonumber\\
&& \,\,\,\,\,
= \int d^4 {\hat x}\int d^4 k \int d^4 p \int d^4 q \,
{\tilde \Phi}(k) ~{\tilde \Psi}(p)
  ~  {\tilde \Upsilon}(q) e^{i(k \oplus p\oplus q)_{\mu}{\hat x}^\mu}
\nonumber\\
&& \,\,\,\,\,
=(2 \pi)^4 \int d^4 k \int d^4 p \int d^4 q \,\,
{\tilde \Phi}(k) ~{\tilde \Psi}(p)
  ~  {\tilde \Upsilon}(q) \, \delta^{(4)}(k \oplus p\oplus q)
\nonumber
\end{eqnarray}
where $\oplus$ is such that
\begin{equation}
(k \oplus p)_0=k_0+p_0
\label{oplus0}
\end{equation}
\begin{equation}
(k \oplus p)_2=k_2+p_2
\label{oplus2}
\end{equation}
\begin{equation}
(k \oplus p)_1 \! = \! \frac{k_2+p_2}{1 \! -e^{\ell(k_2+p_2)}}\!\left[\!
\frac{1 \! - e^{\ell k_2}}{k_2e^{\ell p_2}} k_1
+ \frac{1 \! -e^{\ell p_2}}{p_2} p_1  \! \right]
\label{oplus1}
\end{equation}
This result is rooted in one of the most studied aspects of such noncommutative spacetimes, which is
their ``generalized star product"\ttt\cite{majidLectNotes,lukiestar,alexstar}.  This is essentially a characterization of the properties of
products of exponentials induced by rules of noncommutativity
of type (\ref{kappaspace}). Specifically, one easily arrives at (\ref{soccerball}) (with $\oplus$ such that, in particular,
(\ref{oplus1}) holds) by just observing that from the defining commutator
(\ref{kappaspace})
it follows that\footnote{Eq.\ttt(\ref{star}) is a particular
example of application of
the Baker-Campbell-Hausdorff formula for products of
exponentials of noncommuting variables. In general
the Baker-Campbell-Hausdorff  formula
involves an infinite series of
nested commutators, but the case of
noncommutativity (\ref{kappaspace})
is one of the cases for which the series of
nested commutators
can be resummed explicitly\ttt\cite{lukiestar,alexstar}.}\ttt\cite{lukiestar,alexstar}
\begin{eqnarray}
&&\!\!\log\left[\exp\left({i k_2 \hat{x}_2+i k_1 \hat{x}_1}\right)
\exp\left({ i p_2 \hat{x}_2+i p_1 \hat{x}_1 }\right)\right]=
\label{star}\\
&& =i \hat{x}_2 (p_2+k_2) +i \hat{x}_1
\frac{k_2+p_2}{1 \! -e^{\ell(k_2+p_2)}}\!\left(\!
\frac{1 \! - e^{\ell k_2}}{k_2e^{ \ell p_2}} k_1
+ \frac{1 \! -e^{\ell p_2}}{p_2} p_1  \!  \right)
\nonumber
\end{eqnarray}
The so-called soccer-ball problem concerns the acceptability of
laws of composition of type (\ref{oplus1}). Since one assumes
that the deformation scale $\ell$ is of the order of the inverse of
the Planck scale, applying (\ref{oplus1})
to microscopic/fundamental particles has no sizable consequences:
of course (\ref{oplus1}) gives us back
to good approximation $(k \oplus p)_1 \simeq k_1 + p_1$
whenever $|\ell k_2| \ll 1$ and $|\ell p_2| \ll 1$.
But if a law of composition such as  (\ref{oplus1})
should be used also when we add very many microparticle momenta
in obtaining the total momentum of a multiparticle system (such as a soccer ball) then the
final result could be pathological\ttt\cite{lukiesoccer,maggioresoccer,gacsoccer,kowasoccer,girellisoccer,jacosoccer,sabisoccer,mignemisoccer,magpasoccer,gacfkssoccerRL,sabisoccerRL,gacfkssoccerRL2} even when each microparticle in the system has momentum much smaller than $1/\ell$.

\section{Sum of momenta from translational invariance}
As clarified in the brief review of known results given
in the previous
section, a nonlinear law of composition of momenta arises
in characterizations of locality, as a direct consequence
of the form of some star products. My main point here is that
a different law of composition of momenta is produced by the analysis of translational invariance, and it is this other law of composition of momenta
which is relevant for the characterization of the total momentum
of a multi-particle system. Here too I shall just use known facts
about the peculiarities of translation transformations in
certain noncommutative spacetimes, but exploit them for obtaining
results that had not been derived before, indeed results relevant
for the description of the total momentum
of a multi-particle system.

A first hint that translation transformations should be
modified\ttt\cite{lukieAnnPhys,gaclukieIJMP,jurekphasespace}
 in certain noncommutative spacetimes comes from
noticing that (\ref{kappaspace})
is incompatible with
the standard Heisenberg relations $[p_j,x_k] = i \delta_{jk}$.
Indeed, if one adopts (\ref{kappaspace})
and $[p_j,x_k] = i \delta_{jk}$ one then easily finds that some Jacobi identities
are not satisfied. The relevant Jacobi identities are satisfied
if one allows for a modification of the Heisenberg relations
which balances for the noncommutativity of the
coordinates:
\begin{eqnarray}
&&[p_1,x_1]= i \,\, , \,\,\,\,
 [p_1,x_2]=0 \,\, , \,\,\,\,
[p_2,x_2]= i \,\, , \,\,\,\,\label{kappaphspacetriv}\\
&&[p_1,x_2]=  -i \ell p_1 \,\, , \,\,\,\,
\label{kappaphspace}
\end{eqnarray}
One easily finds that combining
(\ref{kappaspace}),
(\ref{kappaphspacetriv}) and
(\ref{kappaphspace}) all Jacobi identities are satisfied\ttt\cite{lukieAnnPhys,gaclukieIJMP,jurekphasespace}.

Additional intuition for these nonstandard properties of the momenta $p_j$ comes from actually looking at which formulation of translation transformations preserves the form of the noncommutativity of coordinates (\ref{kappaspace}).
 Evidently the standard description
$$x_2 \rightarrow x'_2=x_2+a_2\,\, , \,\,\,\,\,
x_1 \rightarrow x'_1=x_1+a_1$$
is not a symmetry of (\ref{kappaspace}):
\begin{equation}
[x'_1,x'_2]=[x_1+a_1,x_2+a_2]=i \ell x_1 =i \ell(x'_1-a_1)
\label{wrongtrasl}
\end{equation}
Unsurprisingly what does work is the description of translation transformations using as generators the $p_j$ of
(\ref{kappaphspacetriv})-(\ref{kappaphspace}), which as stressed above satisfy
the Jacobi-identity criterion. These deformed translation transformations  take the form
\begin{eqnarray}
 &&x'_1=x_1- i a_1[p_1,x_1]- i a_2[p_2,x_1]
=x_1 + a_1\,\, ,
\nonumber\\
 &&x'_2=x_2- i a_1[p_1,x_2]- i a_2[p_2,x_2]
=x_2 + a_2 - \ell a_1 p_1\,\, \,\,\,\, \,\,\,\, \,\,\,\, \,\,\,\,
\label{traslok}
\end{eqnarray}
 and indeed are symmetries of the commutation rules (\ref{kappaspace}):
 \begin{eqnarray}
[x'_1,x'_2]&=&[x_1+a_1,x_2+a_2-\ell a_1p_1]=\nonumber\\
&=& i \ell x_1-\ell a_1[x_1,p_1] =i \ell(x_1+a_1)=i \ell x'_1
\,\,\,\,\,\, \,\,\,\,\,\, \,\,\,\,\,\, \label{traslokcommut}
\end{eqnarray}
All this about translation transformations in certain noncommutative spacetimes is well known (see, {\it e.g.}, Refs.\ttt\cite{lukieAnnPhys,gaclukieIJMP,jurekphasespace}).
The part which I am here going to contribute is to show how this
is relevant for the mentioned
much-debated issue about the total momentum of
a multi-particle system. My starting point is that  in order
for us to be able to even contemplate the total momentum of a multiparticle
system we must be dealing with a case where translational invariance is ensured:
total momentum is  the conserved charge for a translationally
invariant multi-particle system. Surely the introduction of
translationally
invariant multi-particle systems must involve some subtleties
due to the noncommutativity of coordinates,
and these subtleties are directly connected
to the new properties of translation
transformations (\ref{kappaphspace}),
but they are not directly connected to the properties of the
star product (\ref{star})
 and the associated law of composition of momenta (\ref{oplus1}).
For my purposes, also considering the heated debate that precedes
this contribution of mine,  it is best to show the implications of
this point very simply and explicitly, focusing on a system of
two particles interacting via an harmonic potential.

I start by noticing that
 evidently one does not achieve translational
invariance through a description of the form
\begin{eqnarray}
{\cal H}_{non-transl}=&&\frac{(p_1^A)^2}{2m}+\frac{(p_2^A)^2}{2m}+\frac{(p_1^B)^2}{2m}
+\frac{(p_2^B)^2}{2m}+\nonumber\\
+&&\frac{1}{2}\rho[(x_1^A-x_1^B)^2+(x_2^A-x_2^B)^2]
\end{eqnarray}
where indices $A$ and $B$ label the two particles involved in the
interaction via the harmonic potential.
As stressed above translation transformations consistent with
the coordinate noncommutativity
(\ref{kappaspace})
must be such that
(see (\ref{traslok})) $x_1 \rightarrow x_1 + a_1$
and $x_2 \rightarrow x_2 + a_2 - \ell a_1 p_1$, and as a result
by writing the harmonic potential
with $(x_1^A-x_1^B)^2+(x_2^A-x_2^B)^2$ one does not achieve
translational invariance.

One does get translational invariance by adopting instead
\begin{eqnarray}
{\cal H}=&&\frac{(p_1^A)^2}{2m}+\frac{(p_2^A)^2}{2m}+\frac{(p_1^B)^2}{2m}
+\frac{(p_2^B)^2}{2m}+ \label{harmotrasl}\\
+&&\frac{1}{2}\rho[(x_1^A-x_1^B)^2+(x_2^A
+\ell x_1^A p_1^A-x_2^B - \ell x_1^B p_1^B)^2]
\nonumber
\end{eqnarray}
This is trivially invariant under translations  generated by $p_2$, which simply produce $x_1 \rightarrow x_1$
and $x_2 \rightarrow x_2 +a_2$. And it is also invariant under translations
generated by $p_1$, since they produce $x_1 \rightarrow x_1 + a_1$
and $x_2 \rightarrow x_2 - \ell a_1p_1$,
so that $x_2 + \ell x_1 p_1$ is left unchanged:
$$x_2
+\ell x_1 p_1 \rightarrow x_2
- \ell a_1p_1+\ell( x_1+a_{1} )p_1=x_2
+\ell x_1 p_1$$
It is interesting for my purposes to see which conserved charge
is associated with this invariance under translations of
the hamiltonian $\cal H$. This conserved charge will describe the
total momentum of the two-particle system governed by $\cal H$, {\it i.e.}
the center-of-mass momentum. It is easy to see that this conserved
charge is just the standard $\vec{p}^A+\vec{p}^B$. For the second
component
one trivially finds that indeed
$$[{p}_2^A+{p}_2^B,{\cal H}]=0$$
and the same result also applies to the first component:
\begin{eqnarray}
[{p}_1^A+{p}_1^B,{\cal H}]&& \propto [{p}_1^A+{p}_1^B,(x_1^A-x_1^B)^2]+\nonumber\\
&& \,\, \,\, +[{p}_1^A+{p}_1^B,(x_2^A
+\ell x_1^A p_1^A-x_2^B - \ell x_1^B p_1^B)^2]=\nonumber\\
&& =[{p}_1^A+{p}_1^B,(x_2^A
+\ell x_1^A p_1^A-x_2^B - \ell x_1^B p_1^B)^2]\propto \nonumber\\
&&\propto [{p}_1^A+{p}_1^B,x_2^A
+\ell x_1^A p_1^A-x_2^B - \ell x_1^B p_1^B]\nonumber\\
&& =-i \ell p_1^A + i \ell p_1^A +i \ell p_1^B - i \ell p_1^B
=0
\label{mainresult}
\end{eqnarray}
where the only non-trivial observation I\ have used is that
 (\ref{kappaspace}) leads to $[p_1,x_2
+\ell x_1 p_1] =-i \ell p_1 + i \ell p_1 =0 $.

The result (\ref{mainresult}) shows that indeed $\vec{p}^A+\vec{p}^B$
\uline{is the momentum
of the center of mass} of my translationally-invariant two-particle
system, {\it i.e.} it is the total momentum of the system.

The concerns about total momentum that had been voiced
in discussions of the Planck-scale soccer-ball problem
were rooted in the different sum of momenta
relevant for locality, the $\oplus$ sum discussed in the previous
section. It was feared that one should obtain the total momentum
by combining single-particle momenta with the
nonlinear $\oplus$ sum. The result (\ref{mainresult}) shows that
this expectation was incorrect. One can also directly
verify that indeed $\vec{p}^A \oplus \vec{p}^B$ \uline{is not
a conserved charge} for my translationally-invariant two-particle
system, and specifically, taking into
account (\ref{oplus1}), one finds that
$$[(\vec{p}^A \oplus \vec{p}^B)_1,{\cal H}] \neq 0$$

\section{Implications and outlook}
At least within the framework of spacetime noncommutativity I\ have shown that there is no ``soccer-ball problem", and I\ am confident that analogous results will emerge in other formalisms with nonlinearities in momentum space. This should mature gradually as the understanding of other formalisms matures to the point of appreciating the differences between  the law of composition of momenta obtained by enforcing some form of locality and the law of addition of momenta used for building up the total momentum of a multi-particle system.

As usual in physics, attempts to generalize a theory also help us understand the theory itself: the analysis I\ here reported makes us appreciate
how our current theories are built on a non-trivial correspondence between
the momentum-space manifestations of locality and translational invariance.
This can be viewed from a different perspective by reconsidering the
fact that in Galilean relativity all laws of composition of
momenta and velocities are linear, and there is a linear relationship between velocity and momentum. Within Galilean-relativistic theories one could choose to never speak of momentum, and work exclusively in terms of velocities, with apparently a single linear law of composition of velocities. In our current post-Galilean theories, the relationship between momentum and velocity is non-linear and we then manage to appreciate differences between composition laws (in our current theories all laws of composition of momenta remain linear, but velocities are composed non-linearly).

I\ should devote one final remark to the fact that aspects of my analysis pertaining to  translational invariance were confined to a first-quantized  system. This came out of necessity since several grey areas remain for the formulation of second quantization with $\kappa$-Minkowski noncomutativity.
As a matter of fact I\ here provided the first ever translationally-invariant
formulation of an interacting theory in $\kappa$-Minkowski. All previous
attempts had been made within quantum field theory, and led to unsatisfactory results, particularly for what concerns
global translational invariance.
 Perhaps the results I\ here reported could provide guidance for
improving on previous attempts at formulating interacting
quantum field theories in $\kappa$-Minkowski. In particular, it might be appropriate
 to make room for some novel notion of ``coincidence of points", a possibility which had not been considered in previous attempts.
 I see a hint pointing in this direction in the structure
of my translationally-invariant harmonic potential: unlike standard Harmonic potentials, the potential in my Eq.\ttt(\ref{harmotrasl}) does not vanish when the coordinates of the particles coincide: the potential in
Eq.\ttt(\ref{harmotrasl}) vanishes
for  $x_1^A=x_1^B$ and $x_2^A=x_2^B$, only if also the momenta
coincide (${p}_1^A={p}_1^B$). This is reminiscent of some results
obtained within the recently-proposed relative-locality framework\ttt\cite{prl},
where the only meaningful notion of ``coincidence" is a phase-space notion
(not a notion that could be formulated exclusively in spacetime). This suggests
that one could perhaps improve upon previous attempts of formulation of interacting
quantum field theories in $\kappa$-Minkowski, by
exploiting quantum-field-theory results being developed\ttt\cite{freidelFT}
for the
relative-locality framework.
\bigskip

\bigskip

\bigskip

{\it This work was
supported in part by a grant from the John Templeton Foundation.}

\end{document}